\documentclass[fleqn,usenatbib]{mnras}

\usepackage[T1]{fontenc}
\usepackage{ae,aecompl}

\usepackage{graphicx}	
\usepackage{amsmath}	
\usepackage{amssymb}	
\usepackage{bm}
\usepackage[dvipsnames]{xcolor}

\usepackage{float}

\title[Glass-like point samples in the estimation of $\xi(s)$]{
Improved two-point correlation function estimates using glass-like distributions
as a reference sample}

\author[D\'avila-Kurb\'an et al.]{Federico D\'avila-Kurb\'an$^{1,2}$\thanks{E-mail: fdavilakurban@unc.edu.ar},
Ariel G. S\'anchez$^{3}$,
Marcelo Lares$^{1,2}$,
Andr\'es N. Ruiz$^{1,2}$
\\
$^{1}$Instituto de Astronom\'{\i}a Te\'orica y Experimental (CCT C\'ordoba, CONICET, UNC), Argentina.\\
$^{2}$Observatorio Astron\'omico de C\'ordoba, Universidad Nacional de C\'ordoba, Argentina.\\
$^{3}$Max-Planck-Institut f\"ur Extraterrestrische Physik, Postfach 1312, Giessenbachstr, D-85741 Garching, Germany
}

\date{Accepted XXX. Received YYY; in original form ZZZ}

\pubyear{2020}

\begin{document}
\label{firstpage}
\pagerange{\pageref{firstpage}--\pageref{lastpage}}
\maketitle

\begin{abstract}
All estimators of the two-point correlation function are based on a random catalogue, a set of points with no intrinsic clustering following the selection function of a survey. High-accuracy estimates require the use of large random catalogues, which imply a high computational cost. We propose to replace the standard random catalogues by glass-like point distributions or \textit{glass catalogues} whose power spectrum $P(k)\propto k^4$ exhibits significantly less power on scales larger than the mean inter-particle separation than a Poisson distribution with the same number of points. We show that these distributions can be obtained by iteratively applying the technique of Zeldovich reconstruction commonly used in studies of baryon acoustic oscillations (BAO). We provide a modified version of the widely used Landy-Szalay estimator of the correlation function adapted to the use of glass catalogues and compare its performance with the results obtained using random samples. Our results show that glass-like samples do not add any bias with respect to the results obtained using Poisson distributions. On scales larger than the mean inter-particle separation of the glass catalogues, the modified estimator leads to a significant reduction of the variance of the Legendre multipoles $\xi_\ell(s)$ with respect to the standard Landy-Szalay results with the same number of points. The size of the glass catalogue required to achieve a given accuracy in the correlation function is significantly smaller than when using random samples. Their use could help to drastically reduce the computational cost of configuration-space clustering analysis of future surveys while maintaining high-accuracy requirements.

\end{abstract}




\section{Introduction}

The analysis of the large--scale spatial distribution of galaxies has been instrumental in shaping
our current understanding of the evolution of the Universe 
\citep[e.g.][]{Davis1983, Feldman1994, Efstathiou2002, Tegmark2004, 
Cole2005, Eisenstein2005, Alam2017, eBOSS2020}.
Given the stochastic nature of this distribution, such analyses require robust statistical tools 
to efficiently extract the cosmological information encoded in galaxy surveys.
The most commonly used tools to characterize the large-scale structure of the Universe
are two-point statistics such as the power spectrum, $P(\bm{k})$, and its Fourier transform, the two-point
correlation function $\xi(\bm{ s})$. 
All estimators of these statistics require a set of points that follow the same
selection function of the survey being considered, i.e.,
the position-dependent probability that an object 
is included in the sample, but have no intrinsic clustering. 
In configuration space, the estimators of $\xi({\bf s})$ quantify the
excess probability of finding a pair of galaxies at a given separation vector ${\bf s}$ 
with respect to such reference distribution, often referred to 
as the ``random catalogue'' \citep[]{Peebles1974,Davis1983, Hamilton1993, Landy1993, Baxter2013, Vargas-Magana2013}.
The covariance and bias of these estimators depend on the size of the random catalogue, 
with a higher density sample resulting in more accurate determinations 
\citep[][]{2000ApJ...535L..13K}. 
On the other hand, processing a large random sample can be computationally expensive.
Thus, the estimation of $\xi({\bf s})$ is usually subject to a compromise between the need for 	low 
bias and variance, and maintaining a reasonable computing cost. 

Finding the right balance between accuracy and cost is particularly important in the 
upcoming age of large-volume galaxy surveys, such as 
the dark energy spectroscopic instrument \citep[DESI,][]{desi_survey}, and
the ESA space mission {\it Euclid} \citep{euclid_survey}. 
This problem is exacerbated by the fact that the analysis of galaxy surveys is often accompanied 
by the measurement of the same clustering statistics in thousands of mock catalogues that 
reproduce the properties of the real sample, each of which should ideally have its own specific 
random points \citep{deMattia2019}.
The application of the Zeldovich reconstruction technique \citep{Eisenstein2007,
Padmanabhan2012} commonly used 
in BAO studies  represents an additional complication, as in this case
the estimates of $\xi(\bm{s})$ require to compute pair counts on two different random catalogues.

\begin{figure*}
    \centering
    \includegraphics[width=\textwidth]{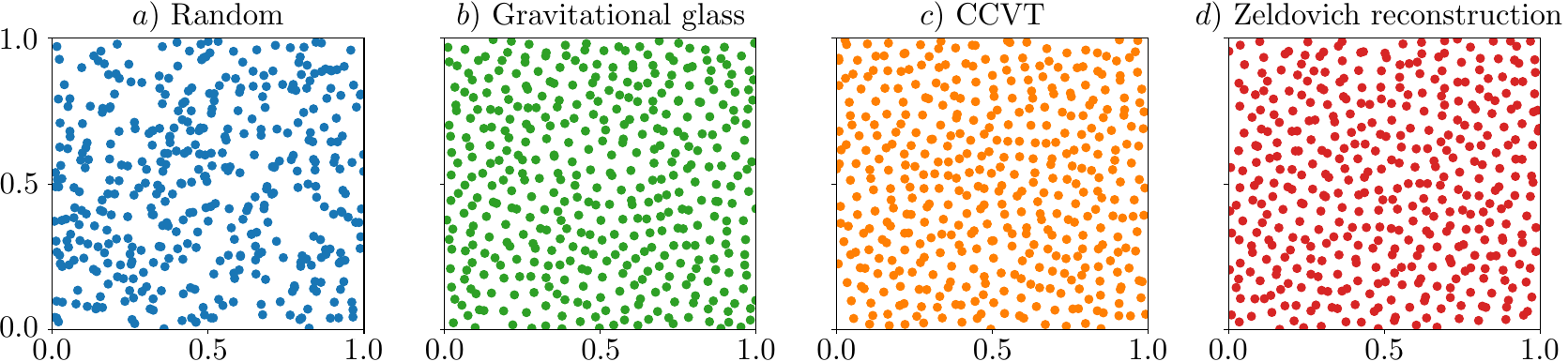}
    \caption{Slices covering 10 per-cent of a cubic box with $16^{3}$ points following a Poisson distribution 
   (panel $a$), a gravitational glass obtained using {\sc gadget}-2 \citep{Springel2005} (panel $b$), a CCVT built using 
   the  code of \citet{Liao2018}  (panel $c$), and a glass-like distribution obtained by iteratively applying 
   Zeldovich reconstruction on an initially random set of points. }
    \label{fig:dist_comparison}
\end{figure*}

The most commonly used estimator of $\xi({\bf s})$ is that of \citet{Landy1993}, hereafter the LS 
estimator, which involves counting the number of data--data pairs in the observed galaxy catalogue 
with some specified separation vector ${\bf s}$, as well as the corresponding number of data--random 
and random--random pairs. As typically the random catalogue is significantly larger than the real sample, 
the random--random component of the estimator dominates the total computing time. 
Several strategies have been proposed to reduce the computational cost of measuring 
$\xi({\bf s})$, such as splitting the random catalogue into smaller sub-samples and averaging the 
pair counts inferred within each of them \citet{Keihanen2019}. However, even following this approach
the computational cost of estimating $\xi({\bf s})$ complicates the analysis of large galaxy samples.

Recently, \citet{Breton2020} proposed a method to estimate pair count terms based on 
analytical expressions that does not rely on the use of a random catalogue. 
This scheme assumes that the selection function 
of the survey can be expressed as the product of an angular footprint, which is described using pixelated 
maps, and a radial distribution, which is estimated from galaxy number counts. For surveys whose 
selection function can be described in this way, the results obtained using this approach are in good 
agreement with those inferred from pair counts based on a random catalogue. However, 
the extension of this method to post-reconstruction measurements, where the displacement 
field inferred from the data must also be applied to the random catalogue, might be non-trivial. 

While the variance of the correlation function is dominated by cosmic variance, in this work we are interested in reducing the error introduced by the random component of correlation estimators.
The impact of the random catalogue on the variance of the correlation function is due to its intrinsic 
density fluctuations, which are characterized by the power spectrum $P(k)=1/\bar{n}$, where $\bar{n}$ 
represents the number density of points. 
Here, we assess whether points following alternative distributions to Poisson, 
whose power spectra exhibit a lower amplitude for the same sample size,
can play the role of the random catalogue and lead to a lower variance in the resulting clustering 
measurements.
Uniform distributions commonly used to generate pre-initial conditions in 
N-body simulations cover the
volume more homogeneously than a Poisson distribution, which makes them a natural 
candidate to replace the standard random catalogues. Examples of such distributions include 
a gravitational glass \citep{White1996} and capacity constrained 
Voronoi tessellations \citep[CCVT,][]{Balzer2009, Liao2018}. As a drawback, the generation of those samples can imply a high computational cost.

Here we propose to use the glass-like distributions obtained by iteratively applying the 
same reconstruction technique used in the context of BAO measurements to a set of initially 
random points.
The small additional computing cost related to the construction of such sample is outweighed by 
the significant improvement in the variance of the estimates of the correlation function with 
respect to the results obtained using Poisson samples of the same size. This also implies that 
the same level of accuracy in the clustering measurements can be achieved with significantly smaller
reference samples, therefore reducing the total computational time.

The outline of this paper is as follows. In Section~\ref{sec:alternative} we review the properties of 
uniform point distributions that can be used as alternatives to the standard random catalogue. We also 
show how a sample with the desired properties can be constructed by applying Zeldovich 
reconstruction to a set of points initially following a Poisson distribution.
In Section~\ref{sec:statest} we review the standard Landy-Szalay estimator and adapt it to the use of 
other uniform distributions. In Section~\ref{sec:zrmethod} we study the differences in the bias and variance 
of the estimates of the Legendre multipoles of the correlation function obtained using random and glass-like 
point distributions as a reference. Finally, Section~\ref{sec:final} presents our main 
conclusions.

\section{Alternatives to the standard random catalogue}
\label{sec:alternative}

\subsection{Homogeneous and isotropic point distributions}
\label{sec:hompondist}

In this section we review a few relevant properties of Poisson and other uniform point distributions and 
discuss the algorithms that can be used to generate them. 

A Poisson distribution is a statistically homogeneous and isotropic distribution characterized by 
a constant power spectrum across all wave numbers, $P(k) =1/\bar{n}$,
where $\bar{n}$ is the number density of points. The power spectrum determines the normalized
variance in the number of points contained in spheres of radius $R$ as \citep[see, e.g.,][]{Gabrielli2002}
\begin{equation}
 \sigma^2(R)=\frac{1}{2\pi^2}\int P(k)W^2(kR)k^2\,{\rm d}k,  
 \label{eq:sigma2}
\end{equation}
where $W(kR)$ is the Fourier transform of the top-hat window function
\begin{equation}
W(y)=3\frac{\sin\left(y\right)-y\cos\left(y\right)}{y^3}.
\end{equation} 
For a Poisson distribution $\sigma^2(R)\propto R^{-3}$.

It is possible to construct sets of points for which $\sigma^2(R)$ decreases faster with $R$ than in the 
Poisson case. 
The fastest possible decay of any distribution is $\sigma^2(R)\propto R^{-4}$ \citep{Gabrielli2002}. 
Samples approaching this limiting behaviour 
are sometimes referred to as \textit{blue noise distributions} 
and
are described by a power--law power spectrum $P(k)\propto k^{4}$,  
which is the minimal large--scale power expected for a discrete stochastic system 
\citep{Peebles1980}. These samples exhibit significantly less power than a Poisson 
distribution for the same number density of points.

An example of sets of points with these properties are the glass-like distributions
commonly used to set up pre-initial conditions of N-body simulations 
\citep{Baugh1993, White1994, Hansen2007, joyce_towards_2009}.
These distributions are obtained by evolving a set of particles that are initially randomly distributed 
under the action of a ``negative'' or repulsive gravitational force until a 
quasi-equilibrium configuration is reached. 
In practice, generating a high-quality gravitational glass with a large number of particles is 
a complex task and 
the associated computational requirements are similar to those of an N-body simulation.

\begin{figure}
 \includegraphics[width=0.95\columnwidth]{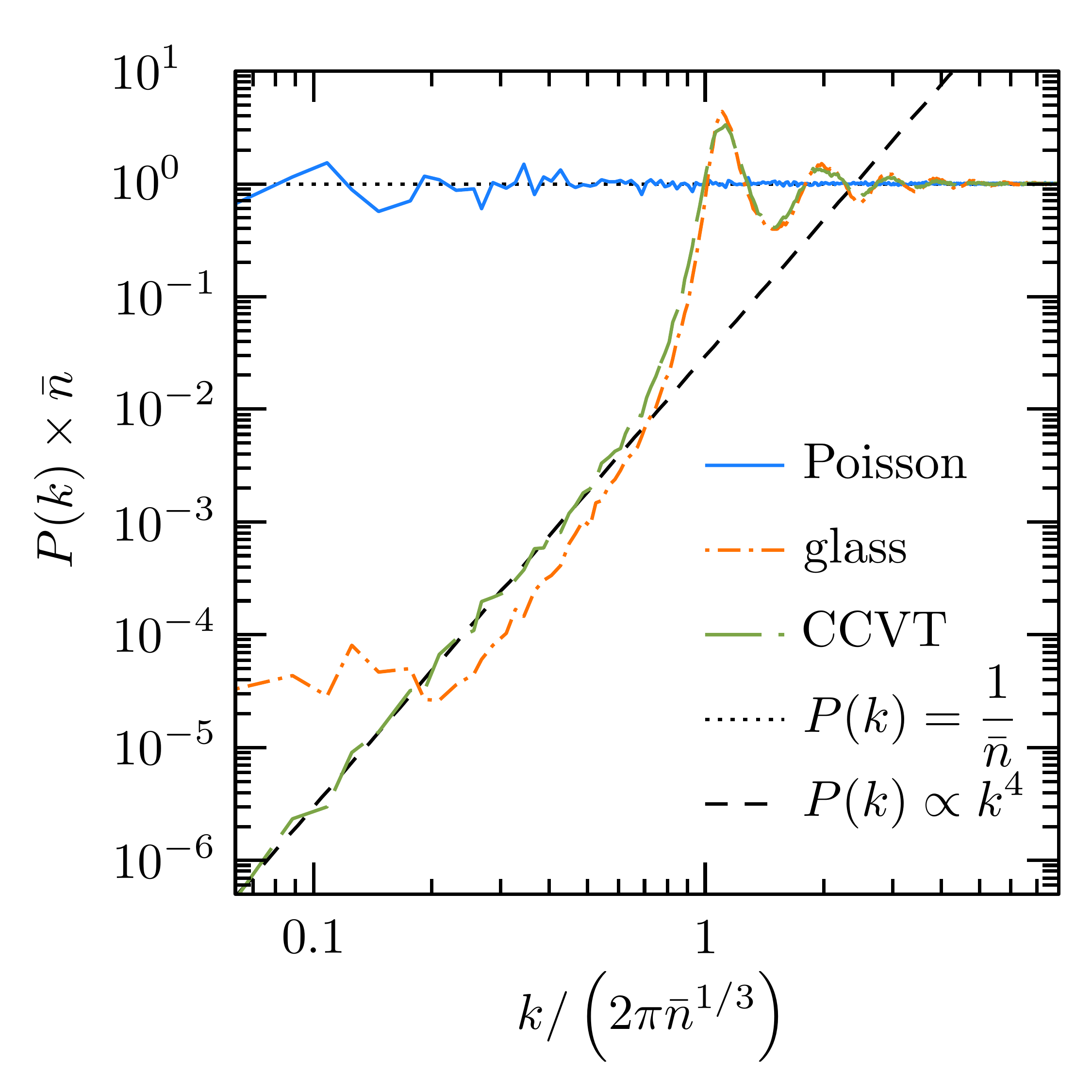}
 \caption{Power spectra of the Poisson (solid), glass (dot-dashed), and CCVT (long-dashed)
 distributions shown in Figure~\ref{fig:dist_comparison}. All power spectra have been rescaled by the mean density 
 of the sample, $\bar{n}$, and wavenumbers by the Fourier-space equivalent of the 
 mean inter-particle separation,  $2\pi/\bar{n}^{-1/3}$.
 The Poisson sample exhibits a roughly constant power over all scales $P(k)=1/\bar{n}$ (dotted). 
 Glass and CCVT distributions follow the  $P(k) \propto k^{4}$ relation expected for 
blue noise distributions (short-dashed). }
 \label{fig:pk}
\end{figure}

A distribution with similar characteristics can be obtained by 
constructing CCVT samples \citep{Balzer2009}, defined by the 
condition that all points are located at the  geometric centre
of their respective Voronoi cells, and that all cells have approximately the same volume.
These highly uniform and isotropic distributions 
have been recently proposed as an alternative to 
the standard gravitational glass as pre-initial conditions of N-body simulations \citep{Liao2018}.
They are generated by iteratively relaxing an initially random set of points into a configuration
satisfying the CCVT conditions \citep{Balzer2009}, but the computing cost of this algorithm becomes prohibitive 
for a large number of points. 

\begin{figure}
 \includegraphics[width=0.95\columnwidth]{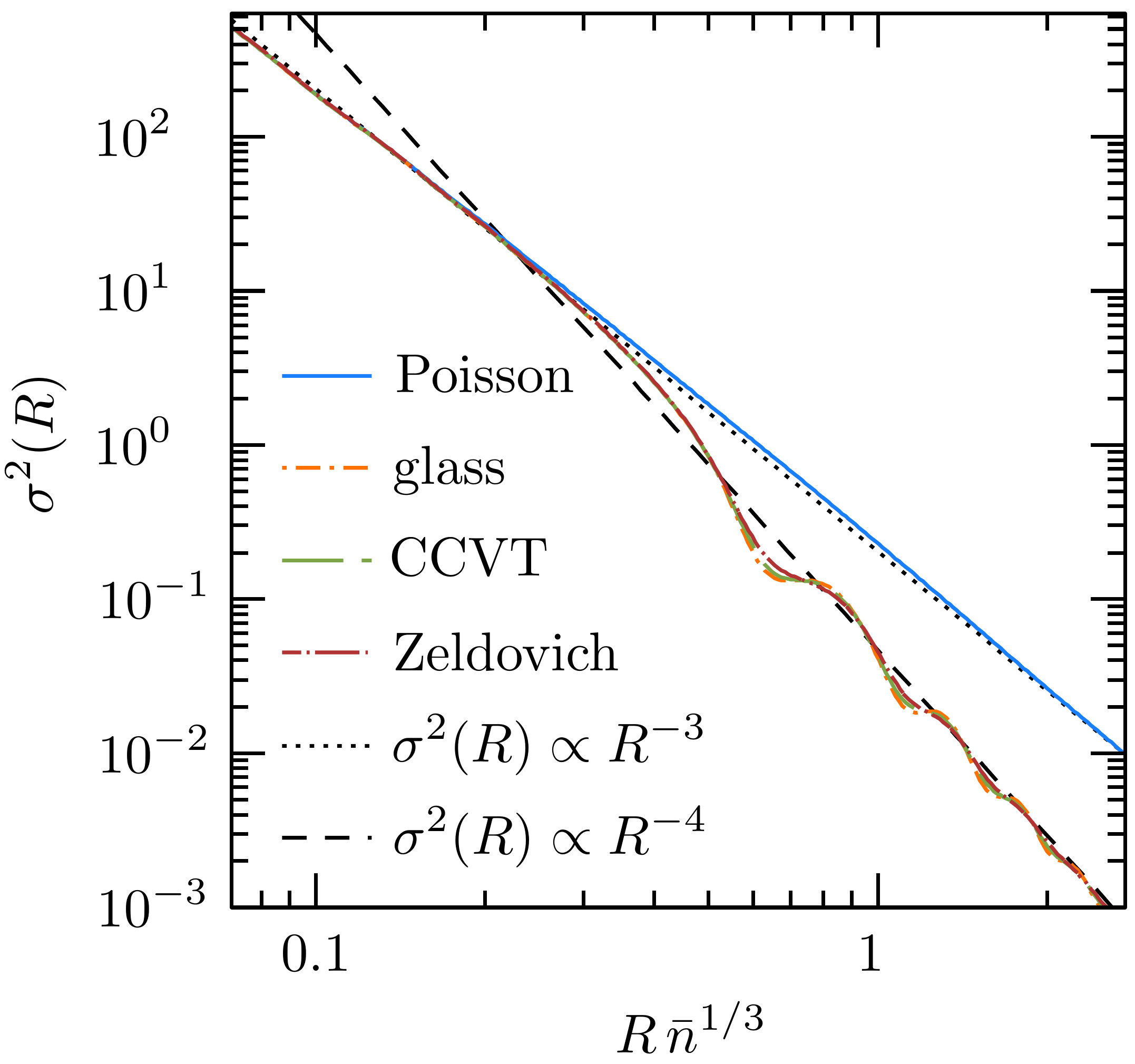}
 \caption{ 
 Normalized variance of the number of points contained in spheres of radius $R$  
 corresponding to the distributions shown in Figure~\ref{fig:dist_comparison},  derived from their power 
spectra using equation~(\ref{eq:sigma2}). While for the Poisson sample $\sigma^2(R)\propto R^{-3}$ 
at all scales, for the blue noise distributions the variance decays as $\sigma^2(R)\propto R^{-4}$ 
for scales larger than the mean inter-particle separation and turns to the behaviour of the random 
sample for smaller scales.}
 \label{fig:integration}
\end{figure}

Figure \ref{fig:dist_comparison} shows a slice covering 10 per-cent of the width of a cubic box 
with $16^{3}$ points following a Poisson distribution (panel $a$), compared against a gravitational glass 
constructed using \textsc{gadget-2} \citep{Springel2005}  and a CCVT distribution obtained using the code of 
\citet{Liao2018} with the same number of points (panels $b$ and $c$, respectively). 
The glass and CCVT distributions cover the volume more homogeneously than the Poisson case, which 
exhibits larger density fluctuations. 
A more quantitative description of the differences between these distributions can be seen in 
Figure~\ref{fig:pk}, which shows the power spectra of the same samples. While the random sample 
exhibits a constant power on all scales, $P(k)=1/\bar{n}$, the Glass and the CCVT distributions 
approach the minimal power spectrum, $P(k)\propto k^{4}$, and turn to the Poissonian behaviour only for scales 
smaller than the mean inter-particle separation, $k\approx 2\pi\bar{n}^{1/3}$. 
Figure~\ref{fig:integration} shows the variance $\sigma^2(R)$ derived from these power 
spectra using equation~(\ref{eq:sigma2}). While for the Poisson distribution $\sigma^2(R)\propto R^{-3}$, 
the variance of the blue noise distributions decays as $\sigma^2(R)\propto R^{-4}$ for scales larger
than the mean inter-particle separation and approaches the behaviour of the random sample 
for smaller scales.

Due to their lower variance, a gravitational glass or a CCVT would be good candidates to 
replace the standard random samples used in the estimation of clustering statistics. 
However, generating these distributions with the number of points required for this task would 
be costly in both computing time and resources.
Although pre-initial conditions of N-body simulations are often built by tiling small periodic glass-like 
distributions to reduce their computing cost, the resulting periodicity of the distribution could have 
undesired effects on the estimation of pair counts.
As we will see in Section~\ref{subsec:zr}, it is possible to construct large sets of points with similar 
properties following a much simpler procedure. 

\subsection{Zeldovich Reconstruction}
\label{subsec:zr}

Lagrangian perturbation theory offers an accurate description of gravitational dynamics for 
small density fluctuations. 
In this section, we show that it can also be used as an alternative to a full N-body force calculation to 
evolve an initially random sample of points into a glass-like state under repulsive gravitational forces
at a significantly smaller computational cost. 

The key quantity in Lagrangian perturbation theory is the displacement field $\bm{\Psi}(\bm{q},t)$,
which maps the initial (Lagrangian) position of a fluid element, $\bm{q}$, to its Eulerian counterpart 
at any given time, $\bm{x}(\bm{q},t)$, as
\begin{equation}
    \bm{x}(\bm{q},t) = \bm{q}+ \bm{\Psi}(\bm{q},t)\,.
\end{equation}
A solution for the displacement field can be found by imposing 
conservation of mass between the Lagrangian and Eulerian coordinate systems, that is
\begin{equation}
    \bar{\rho}d^{3}q = \rho(\bm{x},t)d^{3}x,
\end{equation}
where $\bar{\rho}$ is the mean density and 
$\rho(\bm{x},t)=\bar{\rho}\left(1+\delta\left(\bm{x},t\right)\right)$ represents the Eulerian density 
at position $\bm{x}$ and time $t$. 
Thus, keeping only terms at the linear level, the displacement field is related to the density fluctuations 
in Eulerian space by
\begin{equation}
    \nabla_{\mathbf{q}}\cdot\bm{\Psi}_{(1)}(\bm{q},t) = -\delta_{(1)}(\bm{x},t).
\end{equation}
The subscript (1) indicates that these are first-order terms. Assuming that $\bm{\Psi}$ is an irrotational 
vector field, 
 the solution to this equation can be written in Fourier space as
\begin{equation}
   \bm{\Psi}_{(1)}(\bm{k}) = -\frac{i\bm{k}}{k^{2}}\delta_{(1)}(\bm{k},t).
   \label{eq:displacement}
\end{equation}
This is the solution for the displacement field in first-order Lagrangian perturbation theory and 
corresponds to the standard Zeldovich approximation \citep{Zeldovich1970}. 

Equation~(\ref{eq:displacement})
can be generalized to account for linear redshift-space distortions and galaxy bias and it is the 
basis of the Zeldovich reconstruction technique commonly applied to the analysis of 
 galaxy surveys to enhance the signature of the BAO 
 \citep{Eisenstein2007, Padmanabhan2012, Burden2015}.
The application of the displacement field of equation~(\ref{eq:displacement}) with the opposite sign 
can partially un-do the effects of non-linear gravitational evolution.
The same basic principle can be applied to a set of random points to smooth out its large-scale 
density fluctuations. 

\begin{figure}
\includegraphics[width=0.95\columnwidth]{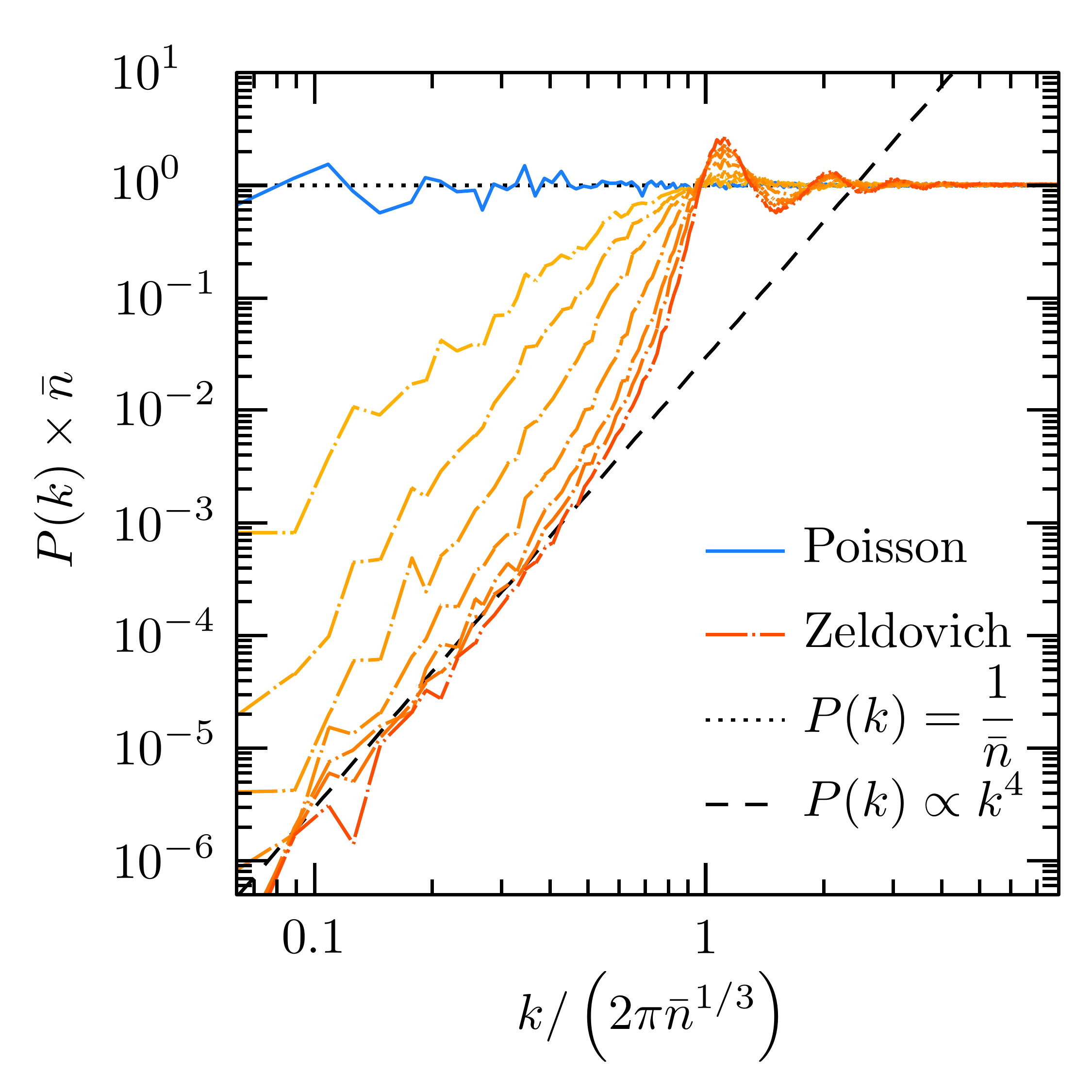}
\caption{Power spectra of the samples obtained by iteratively applying Zeldovich reconstruction to a set of 
random points (dot-dashed lines), shown in the same units as in Fig.~\ref{fig:pk}. 
The results shown correspond to the initial configuration (solid lines) and iterations 1, 2, 4, 10, 20, 30 and 50. 
The power spectrum evolves from pure shot-noise $P(k) =1/\bar{n}$ (dotted lines) to the minimal 
form $P(k)\propto k^{4}$ (dashed lines) for scales larger than the mean inter-particle separation. }

\label{fig:pk_zr_iter}
\end{figure}

Figure~\ref{fig:pk_zr_iter} shows how the iterative application of Zeldovich reconstruction modifies 
the power spectrum of the same set of random points shown in panel $a)$ of Fig.~\ref{fig:dist_comparison}. 
We adapted the publicly available reconstruction code of 
\citet{Bautista2018}\footnote{\url{github.com/julianbautista/eboss\_clustering}}, 
which is based on the Fourier-space algorithm of \citet{Burden2015}. 
We used a fast Fourier transform with a mesh resolution such that the cell size is given by a fourth of 
the mean inter-particle separation. We then estimated densities using a Gaussian kernel with a 
smoothing scale twice the cell size. 
Starting from a constant value across all scales, the amplitude of $P(k)$ decreases towards the 
desired blue noise behaviour, converging onto the minimal-power $P(k) \propto k^{4}$ for 
scales larger than the mean inter-particle separation. 
The the dot -- long-dashed line in Figure~\ref{fig:integration} shows the 
normalized variance $\sigma^2(R)$ corresponding to the point distribution obtained after 50 
iterations, which decays with the steepest slope possible, $\sigma^2(R)\propto R^{-4}$, for 
scales larger than the mean inter-particle separation.
Panel $d)$ of Fig.~\ref{fig:dist_comparison} shows a slice through this final point  
distribution. The points 
cover the volume of the box much more uniformly than the Poisson case. 
The procedure of applying equation~(\ref{eq:displacement}) to a Poisson point distribution is 
significantly simpler and less computationally costly, than generating a glass with a full 
N-body code or building a CCVT distribution. This procedure can also be applied to large samples 
of points.

\section{The estimation of the two-point correlation function}
\label{sec:statest}

\subsection{Revising the Landy-Szalay estimator}
\label{subsec:modifyls}

The most commonly used estimator of the two-point correlation function is 
that of \citet{Landy1993}, which can be computed as 
\begin{equation}
 \xi_{\rm LS}({\bm s}) = \frac{DD({\bm s})-2\,DR({\bm s})+RR({\bm s})}{RR({\bm s})},
\label{eq:ls}
\end{equation}
where $DD(\bm{s})$,  $RR(\bm{s})$, and $DR(\bm{s})$ represent the  
data--data, random--random and data--random pair counts, respectively, for a given 
separation vector $\bm{s}$, 
normalized to the total number of pairs in each case. 

The LS estimator of equation~(\ref{eq:ls}) provides the minimum variance when $|\xi|\ll 1$ and is unbiased 
in the limit of an infinite number of random points, $N_{\rm r} \to \infty$. 
The standard practice for achieving a high accuracy in the measurements of $\xi(\bm{s})$ is to use
a number of random points, $N_{\rm r}$, much larger than the size of the actual data catalogue, $N_{\rm d}$, 
often characterized in terms of the ratio $\alpha\equiv N_{\rm r}/N_{\rm d}$.
Note however that the most relevant quantity to control the bias and variance of the LS 
estimator is the number density of the random catalogue and not the value of $\alpha$.

A larger $N_{\rm r}$ implies an increase in the number of pair counts and therefore of 
the total computational cost.
As discussed in \citet{Keihanen2019}, while the estimation of the 
$RR({\bm s})$ pairs dominates the total computing time, the error 
budget of the estimator is dominated by the term 
$DR({\bm s})$. As a way to speed up the estimation of $\xi(\bm{s})$
without decreasing its accuracy, \citet{Keihanen2019} proposed to split the total set of
random points into $M_{\rm s}$ sub-samples and to approximate the total $RR({\bm s})$ by the 
average of the normalized random-random counts inferred within each of these sub-sets, that is
\begin{equation}
RR({\bm s})\simeq \frac{1}{M_{\rm s}}\sum_{i=1}^{M_{\rm s}}RR_i(\bm{s}),
\label{eq:split}
\end{equation}
where $RR_i(\bm{s})$ represents the results inferred from the $i$-th random sub-sample. 
This approach, dubbed ``split--random'', can reduce the total 
computational time, without affecting the variance or bias of the estimator. However, if 
the analysis must also be performed on a large number of mock catalogues, a large $N_{\rm r}$
would still imply a high computational cost.

We propose to follow an alternative approach to lowering the bias and variance 
of the estimates of $\xi(\bm{s})$ by abandoning the use of random points in 
favour of more uniform distributions such
as the ones described in Sec.~\ref{sec:hompondist}.
Using glass-like point distributions would lower the power spectrum $P(k)$
of the reference sample on larger scales, resulting in a lower 
variance of the pair counts for the same number of points without increasing 
the computational time. In particular, we propose to use the glass-like point samples
obtained after the iterative application of Zeldovich reconstruction to an initially
random distribution as discussed in Sec.~\ref{sec:zrmethod}.

Simply replacing the random sample by a glass in the LS estimator would lead to 
biased clustering measurements. 
As the sample deviates from a Poisson distribution, the position of the points become 
correlated and would yield a biased estimate of  $RR(\bm{s})$. This problem can be avoided 
by using two independent glass-like point distributions or \textit{glass catalogues}, 
$G_1$ and $G_2$,  and modifying the Landy-Szalay estimator as
\begin{equation}
 \xi_{\rm LS,G}(\bm{s}) = \frac{DD(\bm{s})-DG_{1}(\bm{s})-DG_{2}(\bm{s})+G_{1}G_{2}(\bm{s})}{G_{1}G_{2}(\bm{s})}.
\label{eq:mls}
\end{equation}
Apart from $DD(\bm{s})$, all other terms appearing in this expression represent  counts of cross pairs 
between different samples. 
In fact, equation~(\ref{eq:mls}) resembles the generalization of the Landy-Szalay estimator 
commonly used to compute the cross-correlation function between two different data 
samples \citep[e.g.][]{Blake2006}. Appendix \ref{app:estimators} shows generalizations of 
other commonly used estimators of the two- and three-point correlation functions to the use of 
glass catalogues. 

Although the notation of equation~(\ref{eq:mls}) is intended to specify the use of glass-like distributions, 
it can also be implemented with two distinct random catalogues, $R_1$ and $R_2$, that is
\begin{equation}
 \xi_{2R}(\bm{s}) = \frac{DD(\bm{s})-DR_{1}(\bm{s})-DR_{2}(\bm{s})+R_{1}R_{2}(\bm{s})}{R_{1}R_{2}(\bm{s})}.
\label{eq:cross}
\end{equation}
The comparison of the bias and variance of the results obtained using this estimator and equation~(\ref{eq:mls})
can be used to assess the impact of replacing the standard random samples by glass catalogues of the 
same size. Equation~(\ref{eq:cross}) is similar to the split--random method of 
equation~(\ref{eq:split}) with $M_{\rm s}=2$ but using the cross pairs between the two sub-samples to 
infer $RR(\bm{s})$. 
As a reference metric of these different cases, we use the results of the standard LS estimator with a 
single random catalogue containing twice as many points. 

\section{Performance of the estimator}
\label{sec:zrmethod}
\subsection{Methodology}

\begin{figure}
  \centering
  \includegraphics[width=\columnwidth]{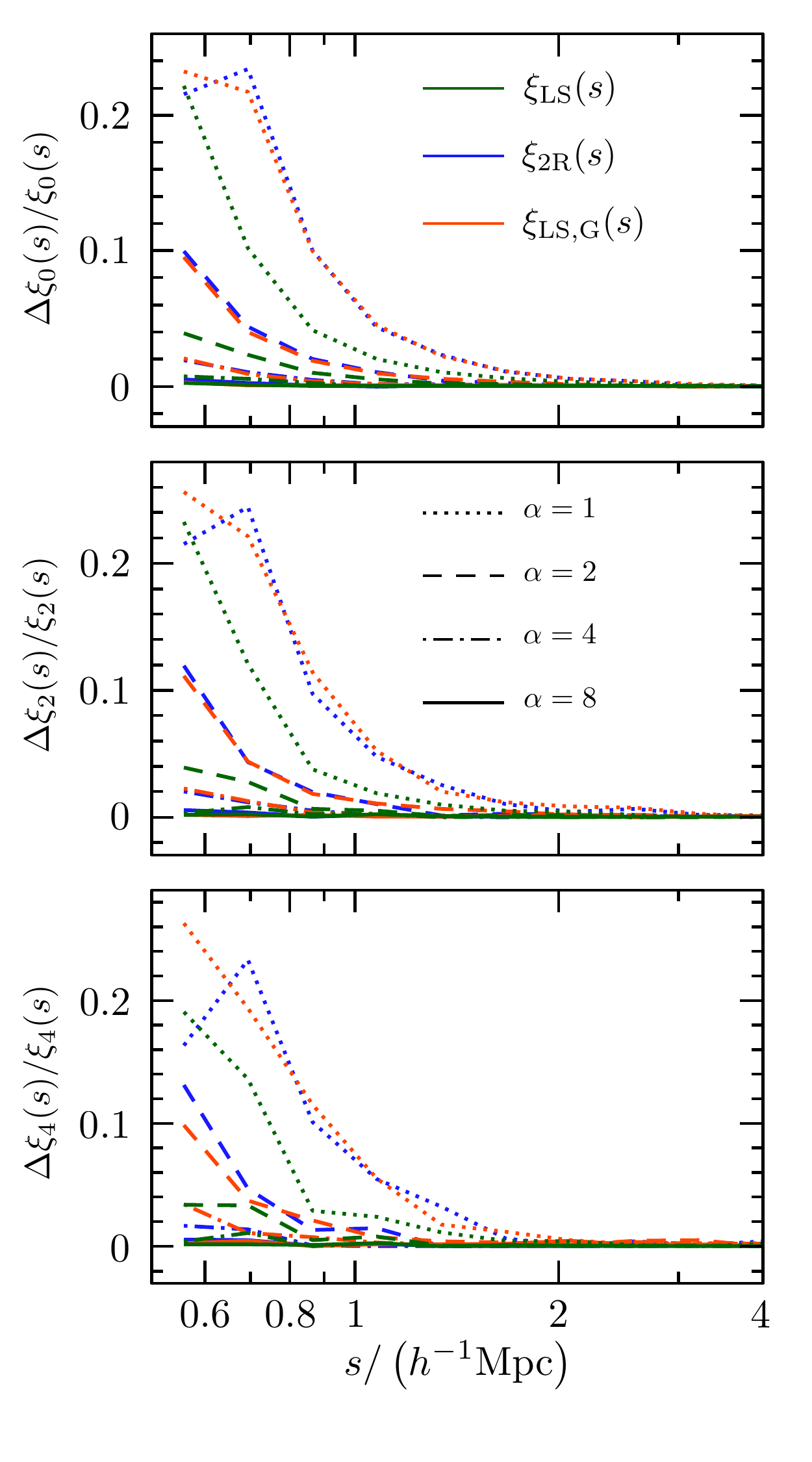}%
\caption{
Relative bias $\Delta\xi_\ell/\xi_\ell(s)$ of the estimates of the monopole (upper panel), quadrupole 
(middle panel) and hexadecapole (lower panel). Line colours indicate the results of the different estimators
described in Section~\ref{subsec:modifyls}, while the line styles represent the total number of points in 
the random and glass catalogues, characterized by values of $\alpha=1,2,4$ and 8.
The estimator of equation~(\ref{eq:mls}) based on glass catalogues show a similar performance 
than the results obtained using random samples of the same size without introducing any additional bias.
}
\label{fig:mean}
\end{figure}

Periodic boundary conditions allow us to 
estimate the two-point correlation function of samples drawn from 
N-body simulations without the use of a random catalogue as
\begin{equation}
\xi(\bm{s})=DD(\bm{s})\frac{V}{\delta V({\bm s})}-1,
\label{eq:xinoran}
\end{equation} 
where $V$ represents the volume of the simulation and $\delta V({\bm s})$ is the volume of the bin 
centred on the pair separation $\bm{s}$ used for the pair counts (e.g. the volume of the spherical 
shell between the radii $s$ and $s+\delta s$ for the angle-averaged correlation function).
The comparison of this exact measurement with the results of the estimators described in 
Sec.~\ref{subsec:modifyls} with random and glass catalogues of different sizes can be used to 
quantify their variance and potential bias.

To this end we used one realization of the \textit{Minerva} N-body simulation suite \citep{Grieb2016, Lippich2019}.
These simulations follow the evolution of the dark matter density field over a cubic box of side length 
$L_{\text{box}} =1.5\,h^{-1}$Gpc with $1000^3$ particles. Each simulation represents a realization of a 
flat $\Lambda$CDM model with physical dark matter and baryon densities $\omega_{\rm c}=0.1154$
and $\omega_{\rm b}= 0.02224$, 
a dimensionless Hubble parameter $h= 0.695$, a scalar spectral index $n_{\rm s}= 0.968$,
and an amplitude of density fluctuations characterized by a linear-theory rms mass fluctuation 
in spheres of radius $12\,{\rm Mpc}$, $\sigma_{12}=0.805$ \citep{Sanchez2020}.
The haloes and subhaloes of these simulations at redshift $z = 0.57$ were populated 
according to a halo occupation distribution (HOD) designed to generate synthetic galaxies with 
clustering properties comparable to the  CMASS sample of the Baryon Oscillation 
Spectroscopic Survey \citep{Dawson2013,Reid2016}.
These HOD catalogues have a mean density of 
$\bar{n}_{\rm d} \approx 4\times 10^{-4} \; h^3 \, \rm{Mpc}^{-3}$.
The impact of redshift-space distortions on galaxy positions was added by taking into account the 
component of their peculiar velocities along one Cartesian axis of the box. 

The full anisotropic correlation function can be described in terms of the modulus of the 
pair separation $s=\left|\bm{s}\right|$ and the cosine of the angle between 
the separation vector $\bm{s}$ and the line
of sight, $\mu$. The information of $\xi(s,\mu)$ can be decomposed into Legendre multipoles, 
$\xi_\ell(s)$, given by 
\begin{equation}
 \xi_\ell(s) = \frac{2\ell+1}{2} \int _{-1} ^1 L_\ell(\mu)\xi(\mu,s) {\rm d} \mu,
 \label{eq:xi_multi}
\end{equation}
where $L_\ell(\mu)$ denotes the Legendre polynomial of order $\ell$. 
We first used the estimators described in Sec.~\ref{subsec:modifyls} to measure $\xi(s,\mu)$
and then used the results to compute Legendre multipoles with $\ell=0,2,4$.
We considered two configurations: one focused on large and intermediate pair separations with 
26 linear bins in the range $20\,\,h^{-1}{\rm Mpc} \leq s \leq 150\,\,h^{-1}{\rm Mpc}$, 
and a second one focussed on small scales with 14 logarithmic bins for pair separations 
$0.5\,\,h^{-1}{\rm Mpc} \leq s \leq 40\,\,h^{-1}{\rm Mpc}$.

We considered random catalogues with mean number densities $\bar{n}$
corresponding to 1, 2, 4 and 8 times that of the HOD sample from Minerva.
For each case, we generated 500 sets of random points and used 
the standard LS estimator to obtain the same number of independent estimates 
of the Legendre multipoles $\xi_{\ell=0,2,4}(s)$ of the Minerva HOD sample. 
We then split each of these random catalogues into two sub-sets with equal 
number of points and used them to measure the same multipoles by applying the 
estimator of equation~(\ref{eq:cross}) and the split-random method of 
equation~(\ref{eq:split}) with $M_{\rm s}=2$. 
These sub-catalogues were then taken as initial points for 
the iterative application of Zeldovich reconstruction as described in Sec.~\ref{sec:zrmethod} 
to generate independent glass catalogues that were used in the estimator of equation~(\ref{eq:mls}). 
The estimates of the multipoles obtained in each case were used to evaluate the 
performance of the different estimators. 
The dispersion of the results recovered using different random catalogues, $\sigma_{\xi_\ell}(s)$, 
quantifies the variance of these estimators, and the mean value of their differences with 
the exact result of equation~(\ref{eq:xinoran}), $\Delta\xi_\ell(s)$, is a measurement of their bias.

\begin{figure}
 \includegraphics[width=\columnwidth]{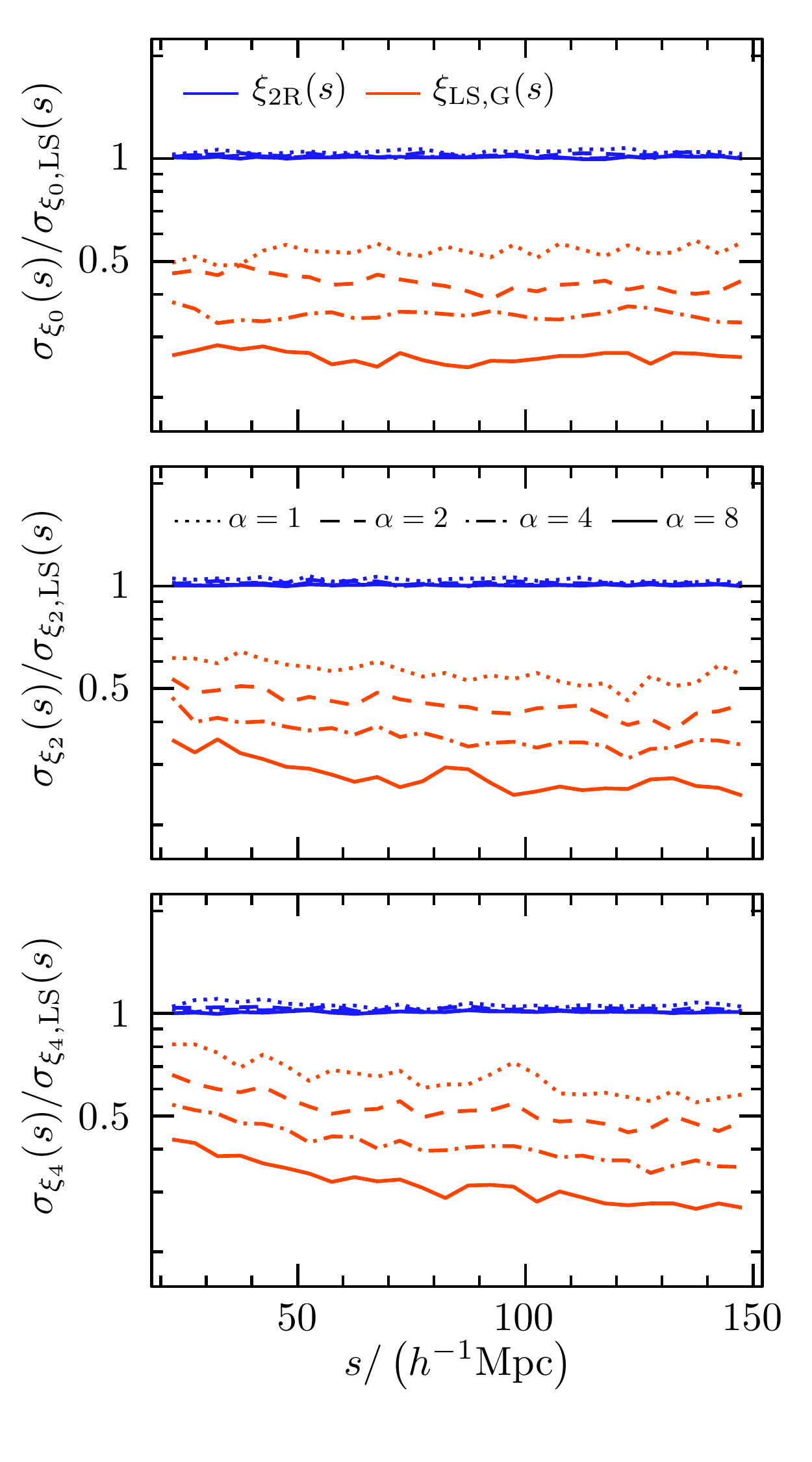}
\caption{
The large-scale standard deviation of the Legendre multipoles $\xi_\ell(s)$ with $\ell=0,2,4$ 
recovered using the estimators of equations~(\ref{eq:mls}) (red) and (\ref{eq:cross}) (blue),
relative to that of the LS case based on the same total number of random points. 
The different line styles indicate the number of points in the random and glass catalogues,
corresponding to $\alpha=1,2,4$ and 8.
While the estimator $\xi_{\rm LS, 2R}(\bm{s})$ has a similar performance to the standard LS, 
the variance obtained using glass catalogues is significantly lower. }
\label{fig:sd_vs_r}
\end{figure}

\begin{figure}
\centering
\includegraphics[width=0.98\columnwidth]{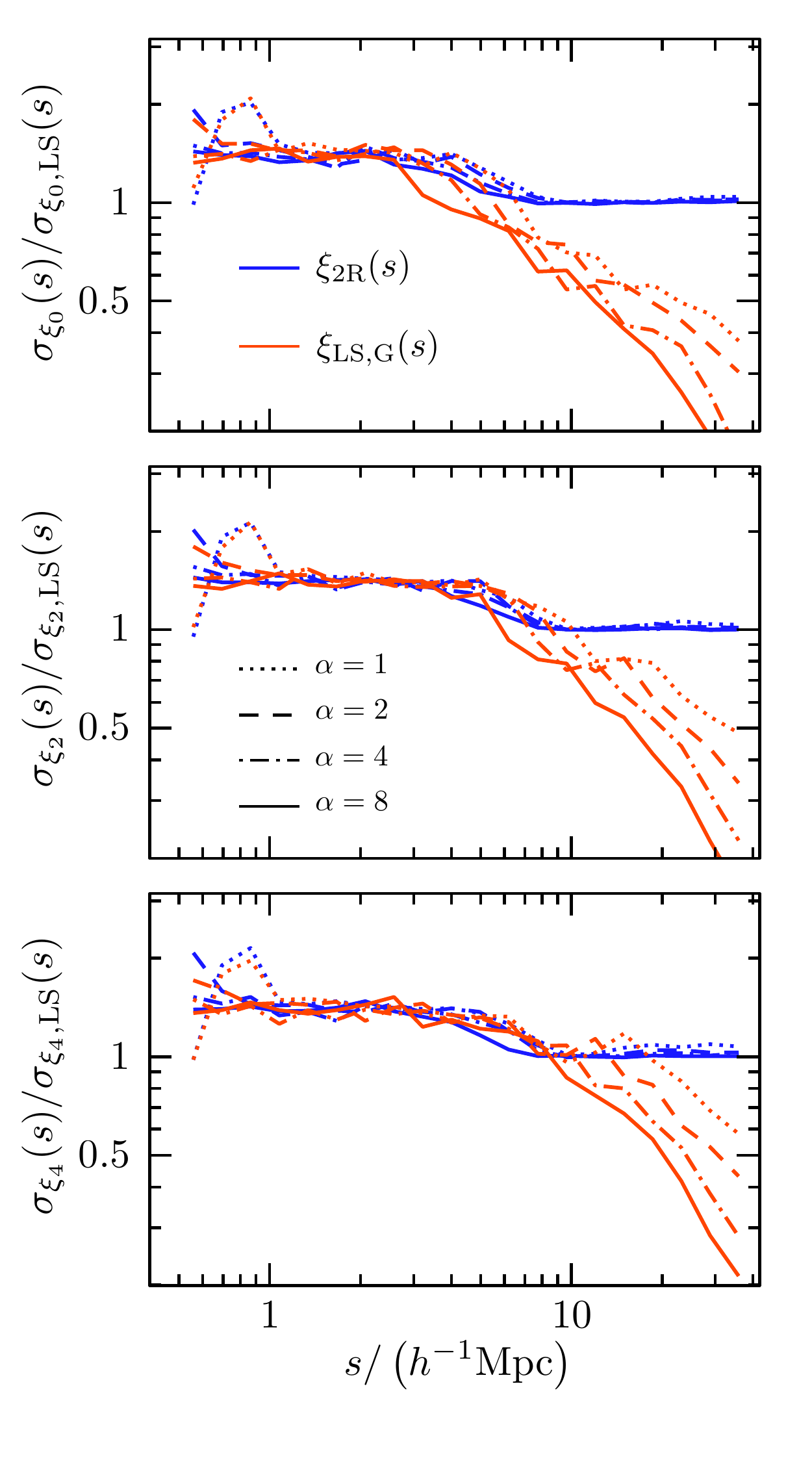}
\caption{
The same as Fig.~\ref{fig:sd_vs_r} but focussing on scales smaller or of the order of the mean
inter-particle separation of the random and glass samples. In this regime, the estimators of 
of equations~(\ref{eq:mls}) (red) and (\ref{eq:cross}) (blue) show a similar performance and 
result in a $\sigma_{\xi_\ell}(s)$ that is $\sim 30$ per-cent higher than that of the standard LS result based
on the same total number of points.}
\label{fig:sd_vs_r_ss}
\end{figure}

\subsection{Bias and variance of the multipoles $\xi_\ell(s)$}

We first focus on the bias of the estimators discussed in Sec.~\ref{subsec:modifyls} by
studying the mean differences between the Legendre multipoles obtained using different 
random and glass catalogues and the exact result of equation~(\ref{eq:xinoran}).
Fig.~\ref{fig:mean} shows the mean relative bias $\Delta\xi_\ell/\xi_\ell(s)$ 
of the estimators, represented by different colours. The
 line styles indicate the total number of points in the random and glass catalogues. 
The panels show separately the results for the multipoles $\ell=0,2,4$. We show only 
results on small pair separations as none of the estimators introduce a significant bias 
on larger scales. On scales significantly smaller than the mean inter-particle 
separation of the random and glass catalogues, the estimator of equation~(\ref{eq:mls}) 
based on glass catalogues shows a similar performance 
than the results obtained using random samples of the same size.

The variance of the multipoles obtained in the different cases on large and intermediate scales 
are shown in Fig.~\ref{fig:sd_vs_r} while those on small scales in Fig.~\ref{fig:sd_vs_r_ss}.
To better judge the performance of the new estimators of equations~(\ref{eq:mls}) and (\ref{eq:cross}), 
we show the results rescaled by the variance of the standard LS case using a random sample with the same 
total number of points. 
As in Fig.\ref{fig:mean}, the different line styles indicate the number of points in the random and glass 
catalogues.
On large scales, the estimator of equation~(\ref{eq:cross}) (blue lines) has essentially the 
same performance as the standard LS case. We have checked that the split-random method with 
$M_{\rm s} = 2$ gives a similar performance. Both methods are then viable alternatives to the standard 
LS estimator on large scales to reduce the total computing time of the $RR$ term without sacrificing 
the accuracy of the measurements. However, on scales smaller than the mean
inter-particle separation the variances of these estimators increase up to a level approximately 
30 per-cent higher than that of the full LS result. 

\begin{figure}
 \includegraphics[width=0.96\columnwidth]{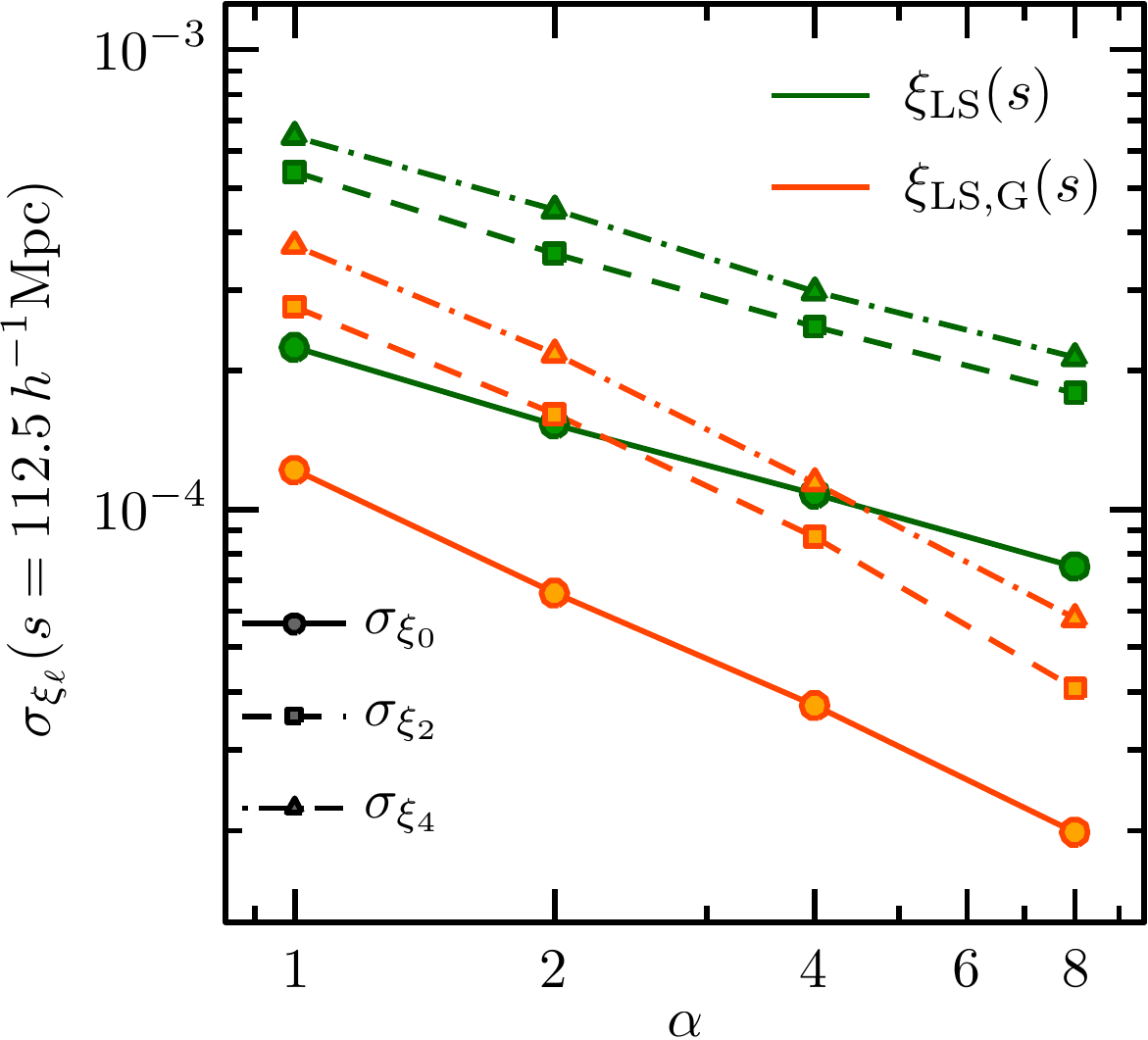}%
 \caption{Standard deviations $\sigma_{\xi_\ell}$ recovered from the LS estimator (green lines) and the 
 modified version of equation~(\ref{eq:mls}) (blue lines) at $s=112.5\,h^{-1}{\rm Mpc}$ as a function of the 
 total number of points in the random and glass catalogues. In both cases, the standard deviation 
 has a power-law dependency on the size of the reference catalogues $N_{\rm r}$, with 
 $\sigma_{\xi_\ell}\propto N_{\rm r}^{-0.53}$ for  $\xi_{\rm LS}(\bm{s})$ and as $\sigma_{\xi_\ell}\propto N_{\rm r}^{-0.86}$ for $\xi_{\rm LS,G}(\bm{s})$. 
 }
 \label{fig:sd_vs_n}
 \end{figure}

Replacing the random catalogues by glass-like samples leads to a markedly different scale 
dependence of $\sigma_{\xi_{\ell}}$ than the standard LS case. 
For scales smaller than the mean inter-particle separation, the variance of the estimator of equation~(\ref{eq:mls})
matches that obtained using two random catalogues. This is expected since, as discussed in 
Section~\ref{sec:hompondist},  on small scales the power spectra of the glass-like 
samples resembles that of a Poisson distribution of the same size.
However, on large scales the variance of $\xi_{\rm LS,G}(\bm{s})$
is significantly smaller 
than that of the LS estimator using a random sample with the same total number of points. 
The improvement with respect to LS becomes more significant with increasing $\alpha$. 
This can be more clearly seen in Fig.~\ref{fig:sd_vs_n}, which shows the variance of the 
Legendre multipoles $\xi_{0,2,4}(s)$ at a scale of $s = 112,5\,h^{-1}{\rm Mpc}$, roughly 
 matching the position of the BAO peak, recovered from the 
LS estimator and equation~(\ref{eq:mls}) as a function of the total number of points in the random 
and glass catalogues. While for $\alpha=1$ the variance of $\xi_{0}(s)$ is $\sim 50$ per-cent smaller 
using glass catalogues, it is only $\sim 25$ per-cent that of the LS for $\alpha=8$. 
The standard deviation of both estimators have a power-law dependency on $N_{\rm r}$ but with different 
slopes. While the variance  decreases as $\sigma_{\xi_\ell}\propto N_{\rm r}^{-0.53}$ for 
$\xi_{\rm LS}(\bm{s})$,  
it goes as $\sigma_{\xi_\ell}\propto N_{\rm r}^{-0.86}$ for $\xi_{\rm LS, G}(\bm{s})$.
This means that a 
given target accuracy of the estimated multipoles can be achieved using glass catalogues that are 
significantly smaller than the random samples that would be required with the standard LS approach.

\section{Conclusions}
\label{sec:final}

We have studied the impact of replacing the random catalogue used in the estimation 
of two-point correlation functions by glass-like point distributions. 
While the power spectrum of these samples resembles that of a Poisson distribution on
scales smaller than the mean inter-particle separation, on larger scales it follows the 
minimal form $P(k)\propto k^4$ and exhibits significantly less power than a
random catalogue with the same number of points. 
In turn, the variance of the counts in spheres of radius $R$ decays as 
$\sigma^2(R)\propto R^{-4}$ as opposed to $\sigma^2(R)\propto R^{-3}$ for the Poisson 
case, which drastically reduces the noise in the pair counts required to estimate $\xi(\bm{s})$. 

We have shown that particle distributions with the desired properties can be generated by 
iteratively applying Zeldovich reconstruction to an initially random set of points. 
This task can take advantage of the fast algorithms that have been developed in 
recent years in the context of BAO reconstruction \citep{Burden2015}. Although we have not 
performed a detailed analysis of the number of iterations required to obtain the 
optimal glass catalogue for correlation function measurements, we have seen that, on the scales 
considered here, the variance of these estimates converges after as few as 5 iterations.
The small additional computing cost associated with the construction of such sample is outweighed by 
the significant improvement in the accuracy of the estimates of the correlation function. 

We have provided a modified version of the LS estimator, adapted to the use of glass-like particle 
distributions (equation~\ref{eq:mls}). This estimator makes use of two independent glass 
catalogues to avoid problems due to the correlation of the points within a single sample for 
scales approaching the mean inter-particle separation. The same estimator can be implemented 
using two different random catalogues (equation~\ref{eq:cross}), offering an ideal benchmark 
to test the advantages of using glass-like particle distributions over Poisson samples. 
As a test case, we used the measurements of the Legendre multipoles $\xi_\ell(s)$ with $\ell=0,2,4$ 
of HOD samples matching the clustering properties of the BOSS CMASS sample obtained 
with different realizations of random and glass catalogues. 

We have found that the large-scale variance of the estimator of equation~(\ref{eq:cross}) using 
two separate random catalogues is similar to that of the standard LS estimator based on a single 
set with the same total number of points. This estimator can then be considered as a simple way to 
reduce the computing cost of measuring $\xi_\ell(s)$ without affecting the accuracy of 
the results, similar to the split-random method of \citet{Keihanen2019}.

Our results show that using glass-like distributions does not add any bias 
with respect to the results obtained using Poisson samples of the same size. 
On scales smaller than the mean inter-particle separation of the reference samples, 
using glass catalogues results in a similar $\sigma_{\xi_\ell}$ as that obtained 
using Poisson distributions. 
However, on larger scales they lead to a significant reduction of the variance of $\xi_\ell(s)$
with respect to the results of the standard LS estimator with the same number of points. 
Furthermore, this reduction becomes larger with increasing $\alpha$.
The size of the glass catalogue required to achieve a given accuracy in the correlation 
function is significantly smaller than when using random samples.  
As the computational cost of these estimators is proportional to the total number of 
pairs, the smaller size of the glass catalogue can represent a significant reduction of the total 
computational time. 
For example, extrapolating the power-law behaviour of $\sigma_{\xi_\ell}$ shown in
Figure~\ref{fig:sd_vs_n} we find that the same variance achieved by the LS estimator using 
random samples of $\alpha=50$ as was done in the final BOSS analyses 
\citep{Alam2017,Sanchez2017}
can be obtained using glass catalogues with $\alpha=5.4$. A more demanding 
estimate based on random catalogues with $\alpha=100$ can be matched using
a glass sample with $\alpha=8.3$.
The total computing time, $t$, of the estimators of equations~(\ref{eq:ls}) and (\ref{eq:mls}) 
is dominated by the $RR$ and $GG$ terms, and hence $t\propto \alpha^2$.
Then, in the previous two examples, the reduction in the computing time obtained by using 
glass catalogues instead of random catalogues amounts to $t_{\rm R}/t_{\rm G}\simeq 86$ and 
$t_{\rm R}/t_{\rm G}\simeq 145$, respectively, representing a drastic reduction of the 
computing resources required for the analysis.
This improvement can be particularly beneficial 
for post-reconstruction BAO measurements, which require pair counts on separate 
random catalogues with and without the application of the displacement field. 

Although we have focused on the LS estimator, Appendix~\ref{app:estimators} 
contains versions of other commonly used estimators of $\xi(\bm{s})$ and $N$-point
correlation functions, adapted to the use of glass catalogues. Estimators based on 
glass-like samples could prove increasingly useful in 
the calculation of high-order statistics, where the computational cost of counting $N$-tuples 
of points could be drastically reduced without compromising variance.
Even though the size of the random catalogue is not the main factor to determine the computational
cost of power spectrum measurements, Fourier-space statistics could also benefit from the lower 
variance of glass catalogues both in the estimation of $P(k)$ itself and the survey window function.
In the coming years, surveys like DESI and Euclid will deliver samples of tens of millions of 
objects over large volumes. The analysis of these catalogues will 
represent a challenge for standard analysis techniques.  
The use of glass catalogues could help to drastically reduce the computational requirements of 
clustering analysis in these surveys and their associated mock catalogues while maintaining the 
high accuracy that they demand.

\section*{Acknowledgements}
This work was partially supported by the Consejo Nacional de
Investigaciones Cient\'{\i}ficas y T\'ecnicas (CONICET, Argentina),
and the Secretar\'{\i}a de Ciencia y Tecnolog\'{\i}a, Universidad Nacional
de C\'ordoba, Argentina.
This project  has  received  financial  support  from  the  European Union 
Horizon 2020 Research and 
 Innovation programme under  the  Marie  Sklodowska-Curie  grant  agreement  number  734374  --  
 project  acronym:  LACEGAL. 
This research has made use of NASA's Astrophysics Data System.
%
%
\section*{Data availability}

The data underlying in this article are available on request to the corresponding author.




\bibliographystyle{mnras}
\bibliography{paper} 

\begin{thebibliography}{}
\makeatletter
\relax
\def\mn@urlcharsother{\let\do\@makeother \do\$\do\&\do\#\do\^\do\_\do\%\do\~}
\def\mn@doi{\begingroup\mn@urlcharsother \@ifnextchar [ {\mn@doi@}
  {\mn@doi@[]}}
\def\mn@doi@[#1]#2{\def\@tempa{#1}\ifx\@tempa\@empty \href
  {http://dx.doi.org/#2} {doi:#2}\else \href {http://dx.doi.org/#2} {#1}\fi
  \endgroup}
\def\mn@eprint#1#2{\mn@eprint@#1:#2::\@nil}
\def\mn@eprint@arXiv#1{\href {http://arxiv.org/abs/#1} {{\tt arXiv:#1}}}
\def\mn@eprint@dblp#1{\href {http://dblp.uni-trier.de/rec/bibtex/#1.xml}
  {dblp:#1}}
\def\mn@eprint@#1:#2:#3:#4\@nil{\def\@tempa {#1}\def\@tempb {#2}\def\@tempc
  {#3}\ifx \@tempc \@empty \let \@tempc \@tempb \let \@tempb \@tempa \fi \ifx
  \@tempb \@empty \def\@tempb {arXiv}\fi \@ifundefined
  {mn@eprint@\@tempb}{\@tempb:\@tempc}{\expandafter \expandafter \csname
  mn@eprint@\@tempb\endcsname \expandafter{\@tempc}}}

\bibitem[\protect\citeauthoryear{Alam et~al.,}{Alam et~al.}{2017}]{Alam2017}
Alam S.,  et~al., 2017, \mn@doi [\mnras] {10.1093/mnras/stx721}, 470, 2617

\bibitem[\protect\citeauthoryear{Balzer, Schl\"{o}mer  \& Deussen}{Balzer
  et~al.}{2009}]{Balzer2009}
Balzer M.,  Schl\"{o}mer T.,   Deussen O.,  2009, in ACM SIGGRAPH 2009 Papers.
  SIGGRAPH '09.
Association for Computing Machinery, New York, NY, USA,
  \mn@doi{10.1145/1576246.1531392}, \url
  {https://doi.org/10.1145/1576246.1531392}

\bibitem[\protect\citeauthoryear{{Baugh} \& {Efstathiou}}{{Baugh} \&
  {Efstathiou}}{1993}]{Baugh1993}
{Baugh} C.~M.,  {Efstathiou} G.,  1993, \mn@doi [\mnras]
  {10.1093/mnras/265.1.145}, \href
  {https://ui.adsabs.harvard.edu/abs/1993MNRAS.265..145B} {265, 145}

\bibitem[\protect\citeauthoryear{{Bautista} et~al.,}{{Bautista}
  et~al.}{2018}]{Bautista2018}
{Bautista} J.~E.,  et~al., 2018, \mn@doi [\apj] {10.3847/1538-4357/aacea5},
  \href {https://ui.adsabs.harvard.edu/abs/2018ApJ...863..110B} {863, 110}

\bibitem[\protect\citeauthoryear{Baxter \& Rozo}{Baxter \&
  Rozo}{2013}]{Baxter2013}
Baxter E.~J.,  Rozo E.,  2013, \mn@doi [\apj] {10.1088/0004-637X/779/1/62}, 779

\bibitem[\protect\citeauthoryear{{Blake}, {Pope}, {Scott}  \&
  {Mobasher}}{{Blake} et~al.}{2006}]{Blake2006}
{Blake} C.,  {Pope} A.,  {Scott} D.,   {Mobasher} B.,  2006, \mn@doi [\mnras]
  {10.1111/j.1365-2966.2006.10158.x}, \href
  {https://ui.adsabs.harvard.edu/abs/2006MNRAS.368..732B} {368, 732}

\bibitem[\protect\citeauthoryear{{Breton} \& {de la Torre}}{{Breton} \& {de la
  Torre}}{2020}]{Breton2020}
{Breton} M.-A.,  {de la Torre} S.,  2020, arXiv e-prints, \href
  {https://ui.adsabs.harvard.edu/abs/2020arXiv201002793B} {p. arXiv:2010.02793}

\bibitem[\protect\citeauthoryear{{Burden}, {Percival}  \& {Howlett}}{{Burden}
  et~al.}{2015}]{Burden2015}
{Burden} A.,  {Percival} W.~J.,   {Howlett} C.,  2015, \mn@doi [\mnras]
  {10.1093/mnras/stv1581}, \href
  {https://ui.adsabs.harvard.edu/abs/2015MNRAS.453..456B} {453, 456}

\bibitem[\protect\citeauthoryear{Cole et~al.,}{Cole et~al.}{2005}]{Cole2005}
Cole S.,  et~al., 2005, \mn@doi [\mnras] {10.1111/j.1365-2966.2005.09318.x},
  362, 505

\bibitem[\protect\citeauthoryear{{DESI Collaboration} et~al.,}{{DESI
  Collaboration} et~al.}{2016}]{desi_survey}
{DESI Collaboration} et~al., 2016, arXiv e-prints, \href
  {https://ui.adsabs.harvard.edu/abs/2016arXiv161100036D} {p. arXiv:1611.00036}

\bibitem[\protect\citeauthoryear{{Davis} \& {Peebles}}{{Davis} \&
  {Peebles}}{1983}]{Davis1983}
{Davis} M.,  {Peebles} P.~J.~E.,  1983, \mn@doi [\apj] {10.1086/160884}, \href
  {https://ui.adsabs.harvard.edu/abs/1983ApJ...267..465D} {267, 465}

\bibitem[\protect\citeauthoryear{{Dawson} et~al.,}{{Dawson}
  et~al.}{2013}]{Dawson2013}
{Dawson} K.~S.,  et~al., 2013, \mn@doi [\aj] {10.1088/0004-6256/145/1/10},
  \href {https://ui.adsabs.harvard.edu/abs/2013AJ....145...10D} {145, 10}

\bibitem[\protect\citeauthoryear{{Efstathiou} et~al.,}{{Efstathiou}
  et~al.}{2002}]{Efstathiou2002}
{Efstathiou} G.,  et~al., 2002, \mn@doi [\mnras]
  {10.1046/j.1365-8711.2002.05215.x}, \href
  {https://ui.adsabs.harvard.edu/abs/2002MNRAS.330L..29E} {330, L29}

\bibitem[\protect\citeauthoryear{Eisenstein et~al.,}{Eisenstein
  et~al.}{2005}]{Eisenstein2005}
Eisenstein D.~J.,  et~al., 2005, \mn@doi [\apj] {10.1086/466512}, 633, 560

\bibitem[\protect\citeauthoryear{{Eisenstein}, {Seo}, {Sirko}  \&
  {Spergel}}{{Eisenstein} et~al.}{2007}]{Eisenstein2007}
{Eisenstein} D.~J.,  {Seo} H.-J.,  {Sirko} E.,   {Spergel} D.~N.,  2007,
  \mn@doi [\apj] {10.1086/518712}, \href
  {https://ui.adsabs.harvard.edu/abs/2007ApJ...664..675E} {664, 675}

\bibitem[\protect\citeauthoryear{Feldman, Kaiser  \& Peacock}{Feldman
  et~al.}{1994}]{Feldman1994}
Feldman H.~A.,  Kaiser N.,   Peacock J.~A.,  1994, \mn@doi [\apj]
  {10.1086/174036}, 426, 23

\bibitem[\protect\citeauthoryear{{Gabrielli}, {Joyce}  \& {Sylos
  Labini}}{{Gabrielli} et~al.}{2002}]{Gabrielli2002}
{Gabrielli} A.,  {Joyce} M.,   {Sylos Labini} F.,  2002, \mn@doi [\prd]
  {10.1103/PhysRevD.65.083523}, \href
  {https://ui.adsabs.harvard.edu/abs/2002PhRvD..65h3523G} {65, 083523}

\bibitem[\protect\citeauthoryear{Grieb, S{\'{a}}nchez, Salazar-Albornoz  \&
  Vecchia}{Grieb et~al.}{2016}]{Grieb2016}
Grieb J.~N.,  S{\'{a}}nchez A.~G.,  Salazar-Albornoz S.,   Vecchia C.~D.,
  2016, \mn@doi [\mnras] {10.1093/mnras/stw065}, 457, 1577

\bibitem[\protect\citeauthoryear{Hamilton}{Hamilton}{1993}]{Hamilton1993}
Hamilton A.,  1993, \mn@doi [\apj] {10.1086/173288}, 417, 19

\bibitem[\protect\citeauthoryear{{Hansen}, {Agertz}, {Joyce}, {Stadel}, {Moore}
   \& {Potter}}{{Hansen} et~al.}{2007}]{Hansen2007}
{Hansen} S.~H.,  {Agertz} O.,  {Joyce} M.,  {Stadel} J.,  {Moore} B.,
  {Potter} D.,  2007, \mn@doi [\apj] {10.1086/510477}, \href
  {https://ui.adsabs.harvard.edu/abs/2007ApJ...656..631H} {656, 631}

\bibitem[\protect\citeauthoryear{{Joyce}, {Marcos}  \& {Baertschiger}}{{Joyce}
  et~al.}{2009}]{joyce_towards_2009}
{Joyce} M.,  {Marcos} B.,   {Baertschiger} T.,  2009, \mn@doi [\mnras]
  {10.1111/j.1365-2966.2008.14290.x}, \href
  {https://ui.adsabs.harvard.edu/abs/2009MNRAS.394..751J} {394, 751}

\bibitem[\protect\citeauthoryear{{Keih{\"a}nen} et~al.,}{{Keih{\"a}nen}
  et~al.}{2019}]{Keihanen2019}
{Keih{\"a}nen} E.,  et~al., 2019, \mn@doi [\aap] {10.1051/0004-6361/201935828},
  \href {https://ui.adsabs.harvard.edu/abs/2019A&A...631A..73K} {631, A73}

\bibitem[\protect\citeauthoryear{Kerscher, Szapudi  \& Szalay}{Kerscher
  et~al.}{2000}]{2000ApJ...535L..13K}
Kerscher M.,  Szapudi I.,   Szalay A.~S.,  2000, \mn@doi [\apjl]
  {10.1086/312702}, 535, L13

\bibitem[\protect\citeauthoryear{Landy \& Szalay}{Landy \&
  Szalay}{1993}]{Landy1993}
Landy S.~D.,  Szalay A.~S.,  1993, \mn@doi [\apj] {10.1086/172900}, 412, 64

\bibitem[\protect\citeauthoryear{{Laureijs} et~al.,}{{Laureijs}
  et~al.}{2011}]{euclid_survey}
{Laureijs} R.,  et~al., 2011, arXiv e-prints, \href
  {https://ui.adsabs.harvard.edu/abs/2011arXiv1110.3193L} {p. arXiv:1110.3193}

\bibitem[\protect\citeauthoryear{{Liao}}{{Liao}}{2018}]{Liao2018}
{Liao} S.,  2018, \mn@doi [\mnras] {10.1093/mnras/sty2523}, \href
  {https://ui.adsabs.harvard.edu/abs/2018MNRAS.481.3750L} {481, 3750}

\bibitem[\protect\citeauthoryear{Lippich et~al.,}{Lippich
  et~al.}{2019}]{Lippich2019}
Lippich M.,  et~al., 2019, \mn@doi [\mnras] {10.1093/mnras/sty2757}, 482, 1786

\bibitem[\protect\citeauthoryear{{Padmanabhan}, {Xu}, {Eisenstein}, {Scalzo},
  {Cuesta}, {Mehta}  \& {Kazin}}{{Padmanabhan} et~al.}{2012}]{Padmanabhan2012}
{Padmanabhan} N.,  {Xu} X.,  {Eisenstein} D.~J.,  {Scalzo} R.,  {Cuesta} A.~J.,
   {Mehta} K.~T.,   {Kazin} E.,  2012, \mn@doi [\mnras]
  {10.1111/j.1365-2966.2012.21888.x}, \href
  {https://ui.adsabs.harvard.edu/abs/2012MNRAS.427.2132P} {427, 2132}

\bibitem[\protect\citeauthoryear{Peebles}{Peebles}{1980}]{Peebles1980}
Peebles P.,  1980, {The large-scale structure of the universe}.
Princeton University Press

\bibitem[\protect\citeauthoryear{{Peebles} \& {Hauser}}{{Peebles} \&
  {Hauser}}{1974}]{Peebles1974}
{Peebles} P.~J.~E.,  {Hauser} M.~G.,  1974, \mn@doi [\apjs] {10.1086/190308},
  \href {https://ui.adsabs.harvard.edu/abs/1974ApJS...28...19P} {28, 19}

\bibitem[\protect\citeauthoryear{{Reid} et~al.,}{{Reid}
  et~al.}{2016}]{Reid2016}
{Reid} B.,  et~al., 2016, \mn@doi [\mnras] {10.1093/mnras/stv2382}, \href
  {https://ui.adsabs.harvard.edu/abs/2016MNRAS.455.1553R} {455, 1553}

\bibitem[\protect\citeauthoryear{{S{\'a}nchez}}{{S{\'a}nchez}}{2020}]{Sanchez2020}
{S{\'a}nchez} A.~G.,  2020, \mn@doi [\prd] {10.1103/PhysRevD.102.123511}, \href
  {https://ui.adsabs.harvard.edu/abs/2020PhRvD.102l3511S} {102, 123511}

\bibitem[\protect\citeauthoryear{S{\'{a}}nchez et~al.,}{S{\'{a}}nchez
  et~al.}{2017}]{Sanchez2017}
S{\'{a}}nchez A.~G.,  et~al., 2017, \mn@doi [\mnras] {10.1093/mnras/stw2443},
  464, 1640

\bibitem[\protect\citeauthoryear{{Springel}}{{Springel}}{2005}]{Springel2005}
{Springel} V.,  2005, \mn@doi [\mnras] {10.1111/j.1365-2966.2005.09655.x},
  \href {https://ui.adsabs.harvard.edu/abs/2005MNRAS.364.1105S} {364, 1105}

\bibitem[\protect\citeauthoryear{{Szapudi} \& {Szalay}}{{Szapudi} \&
  {Szalay}}{1998}]{Szapudi1998}
{Szapudi} I.,  {Szalay} A.~S.,  1998, \mn@doi [\apjl] {10.1086/311146}, \href
  {https://ui.adsabs.harvard.edu/abs/1998ApJ...494L..41S} {494, L41}

\bibitem[\protect\citeauthoryear{{Tegmark} et~al.,}{{Tegmark}
  et~al.}{2004}]{Tegmark2004}
{Tegmark} M.,  et~al., 2004, \mn@doi [\apj] {10.1086/382125}, \href
  {https://ui.adsabs.harvard.edu/abs/2004ApJ...606..702T} {606, 702}

\bibitem[\protect\citeauthoryear{{Vargas-Maga{\~n}a}
  et~al.,}{{Vargas-Maga{\~n}a} et~al.}{2013}]{Vargas-Magana2013}
{Vargas-Maga{\~n}a} M.,  et~al., 2013, \mn@doi [\aap]
  {10.1051/0004-6361/201220790}, \href
  {https://ui.adsabs.harvard.edu/abs/2013A&A...554A.131V} {554, A131}

\bibitem[\protect\citeauthoryear{{White}}{{White}}{1994}]{White1994}
{White} S.~D.~M.,  1994, Reviews in Modern Astronomy, \href
  {https://ui.adsabs.harvard.edu/abs/1994RvMA....7..255W} {7, 255}

\bibitem[\protect\citeauthoryear{{White}}{{White}}{1996}]{White1996}
{White} S.~D.~M.,  1996, in {Schaeffer} R.,  {Silk} J.,  {Spiro} M.,
  {Zinn-Justin} J.,  eds, Cosmology and Large Scale Structure. p.~349

\bibitem[\protect\citeauthoryear{{Zel'Dovich}}{{Zel'Dovich}}{1970}]{Zeldovich1970}
{Zel'Dovich} Y.~B.,  1970, \aap, \href
  {https://ui.adsabs.harvard.edu/abs/1970A&A.....5...84Z} {500, 13}

\bibitem[\protect\citeauthoryear{{de Mattia} \& {Ruhlmann-Kleider}}{{de Mattia}
  \& {Ruhlmann-Kleider}}{2019}]{deMattia2019}
{de Mattia} A.,  {Ruhlmann-Kleider} V.,  2019, \mn@doi [\jcap]
  {10.1088/1475-7516/2019/08/036}, \href
  {https://ui.adsabs.harvard.edu/abs/2019JCAP...08..036D} {2019, 036}

\bibitem[\protect\citeauthoryear{{eBOSS Collaboration} et~al.,}{{eBOSS
  Collaboration} et~al.}{2021}]{eBOSS2020}
{eBOSS Collaboration} et~al., 2021, \mn@doi [\prd]
  {10.1103/PhysRevD.103.083533}, \href
  {https://ui.adsabs.harvard.edu/abs/2021PhRvD.103h3533A} {103, 083533}

\makeatother
\end{thebibliography}

\appendix

\section{Additional estimators of the two- and three-point correlation
functions}
\label{app:estimators}

In this section we give alternative versions of the most common estimators of 
two- and $N$-point correlation functions, adapted to the use of glass 
catalogues instead of the standard random distributions.

The simplest estimator of the two-point function is the so-called natural estimator
\citep{Peebles1974}, which can be modified to use two glass catalogues as
\begin{equation}
\xi_{\rm PH,G}(\bm{s}) = \frac{DD(\bm{s})}{G_1G_2(\bm{s})}-1.
\label{eq:natural_glass}
\end{equation}
The estimator of \citet{Davis1983} can be implemented with a single glass
catalogue as
\begin{equation}
\xi_{\rm DP,G}(\bm{s}) = \frac{DD(\bm{s})}{DG(\bm{s})}-1.
\label{eq:dp_glass}
\end{equation}
Another commonly used estimator is that of \citet{Hamilton1993}, which
can be adapted to the case of two glass catalogues as
\begin{equation}
\xi_{\rm H, G}(\bm{s}) = \frac{DD(\bm{s})G_1G_2(\bm{s})}{DG_1(\bm{s})\,DG_2(\bm{s})}-1.
\label{eq:ham_glass}
\end{equation}

The estimation of the two-point correlation function after the application of Zeldovich reconstruction  requires the use of two different 
random catalogues, one following the original selection function, and a second one, denoted by $S$,
that is shifted by applying the same displacement field as to the data $D$. 
The estimator of equation~(\ref{eq:ls}) is then modified as \citep{Padmanabhan2012}
\begin{equation}
 \hat{\xi}_{\rm rec}(\bm{s}) = \frac{DD(\bm{s})-2DS(\bm{s})+SS(\bm{s})}{RR(\bm{s})}.
\end{equation}
Adapting this estimator to the use of glass catalogues requires four different glass-like samples, two
of which follow the original data, $G_{1,2}$, and two in which the displacement field has been applied, 
$G_{{\rm S}1,2}$, leading to
\begin{equation}
 \hat{\xi}_{\rm rec,G}(\bm{s}) = \frac{DD(\bm{s})-DG_{{\rm S}1}(\bm{s})-DG_{{\rm S}2}(\bm{s})+G_{{\rm S}1}G_{{\rm S}2}(\bm{s})}{G_{1}G_{2}(\bm{s})}.
\end{equation}

Estimators of higher-order correlation functions can also be adapted to use 
glass catalogues. In the general notation of \citet{Szapudi1998}, the estimator 
for the $N$-point correlation function can be written as
\begin{equation}
\xi_{N,{\rm G}} = \left(D-G_1\right)\left(D-G_2\right)\cdots\left(D-G_N\right)/G_1G_2\cdots G_N,
\end{equation}
which requires $N$ independent glass catalogues $G_i$.
For the two-point correlation function, this expression corresponds to the modified LS estimator of equation~(\ref{eq:mls}).
For the three-point correlation function, this estimator can be expressed as
\begin{align}
\xi_{\rm 3, G}(s_{12},s_{23},s_{13}) &= \frac{DDD(s_{12},s_{23},s_{13})}{G_1G_2G_3(s_{12},s_{23},s_{13})}  \label{eq:xi3_glass}\\
&- \frac{\sum_{i}DDG_i(s_{12},s_{23},s_{13})}{G_1G_2G_3(s_{12},s_{23},s_{13})}  \nonumber\\
 &+ \frac{\sum_{(i,j) \in \binom{{\mathcal I}}{2}}DG_iG_j(s_{12},s_{23},s_{13})}{G_1G_2G_3(s_{12},s_{23},s_{13})} +1,\nonumber
\end{align}
where ${\mathcal I} = \left\{1,2,3\right\}$, and $\binom{I}{2}$ represents the set of all possible 
combinations of  two elements of ${\mathcal I}$. 
 
Estimators based on glass catalogues could significantly reduce the variance of measurements 
of high-order statistics with respect to the results obtained with a single random catalogue with 
the same  total number of points. Using 
smaller glass catalogues could then significantly reduce the total computational cost of estimating
three-point correlation functions without affecting the accuracy of the measurements. A detailed analysis
of the performance of the estimator of equation~(\ref{eq:xi3_glass}) is left for future work.

\bsp	
\label{lastpage}
\end{document}